\documentclass[aps,preprint]{revtex4}%
\usepackage{amsfonts}%
\usepackage{amsmath}%
\usepackage{amssymb}%
\usepackage{graphicx}

\begin{document}
\preprint{quant-ph/0210086}
\title{Proposal to produce long-lived mesoscopic superpositions through an
atom-driven field interaction{\Large  }}
\author{C. J. Villas-B\^{o}as, F. R. de Paula, R. M. Serra, and M. H. Y. Moussa.}
\affiliation{Departamento de F\'{\i}sica, Universidade Federal de S\~{a}o Carlos, PO Box
676, S\~{a}o Carlos, 13565-905, SP, Brazil.}

\begin{abstract}
We present a proposal for the production of longer-lived mesoscopic
superpositions which relies on two requirements: parametric amplification and
squeezed vacuum reservoir for cavity-field states. Our proposal involves the
interaction of a two-level atom with a cavity field which is simultaneously
subjected to amplification processes.


\textbf{Journal-ref:} \ J. Opt. B: Quantum Semiclass. Opt. \textbf{5}, 391 (2003)

\end{abstract}
\maketitle

The mastery of techniques for preparing cavity-field states through atom-field
interaction in cavity quantum electrodynamics (QED) is crucial to many useful
applications in quantum optics. As high-Q cavities have permitted the
preparation of cavity-field superpositions of the form $\left\vert
\Psi\right\rangle =$ $\left(  \left\vert \alpha\operatorname{e}^{i\phi
}\right\rangle +\left\vert \alpha\operatorname{e}^{-i\phi}\right\rangle
\right)  /\sqrt{2}$, with mean number of oscillator quanta $\left\vert
\alpha\right\vert ^{2}\approx10$, mesoscopic quantum coherence in cavity QED
has been investigated: the progressive decoherence of the superposition
$\left\vert \Psi\right\rangle $, involving radiation fields with classically
distinct phases, was observed through atom-field interaction \cite{Brune} and
the reversible decoherence of such a cavity-field state has been conjectured
\cite{Raimond}. In this letter we present a proposal for the achievement of
long-lived mesoscopic superposition states in cavity QED which relies on two
basic requirements: parametric amplification and an engineered squeezed-vacuum
reservoir for cavity-field states. Our proposal considers the dispersive
interaction of a two-level atom with a cavity field which is simultaneously
under amplification processes. The parametric amplification is employed to
achieve the required high degrees of squeezing and excitation of what we
actually want to be a mesoscopic superposition state. Such long-lived
squeezed-mesoscopic state, under the action of a likewise squeezed reservoir,
exhibit a decoherence time order of magnitudes longer than those for
non-squeezed cavity-field states subjected to the influence of $i)$ a squeezed
reservoir and $ii)$ a non-squeezed reservoir.

\textit{Atom-driven field interaction}: The proposed configuration for
engineering driven-cavity-field states, based on the scheme by Brune
\textit{et al}. \cite{Brune1}, consists of a two-level Rydberg atom which
crosses a Ramsey-type arrangement, i.e., a high-Q micromaser cavity $C$
located between two Ramsey zones. After interacting with this arrangement, the
atom is counted by ionization chambers, projecting the cavity-field state in
$C$. The transition $|2\rangle\rightarrow|1\rangle$ of the two-level atom
(excited $|2\rangle$ and ground state $|1\rangle$) is far from resonant with
the cavity mode frequency, allowing for a dispersive atom-field interaction.
In addition to the interaction with the two-level atom, the cavity mode is
simultaneously submitted to linear and parametric amplifications so that the
Hamiltonian of our model is given by ($\hbar=1$)%

\begin{equation}
H=\omega a^{\dagger}a+\frac{\omega_{0}}{2}\sigma_{z}+\chi a^{\dagger}%
a\sigma_{z}+\zeta(t)a^{\dagger^{2}}+\zeta^{\ast}(t)a^{^{2}}+\xi(t)a^{\dagger
}+\xi^{\ast}(t)a\mathrm{{,}}\label{Eq1}%
\end{equation}
where $\sigma_{z}=|2\rangle\langle2|-|1\rangle\langle1|$, $a$ and $a^{\dagger
}$ are, respectively, the creation and annihilation operators for the cavity
mode of frequency $\omega$ which lies between the two atomic energy levels,
which are separated by $\omega_{0}$, such that the detuning $\delta
=|\omega-\omega_{0}|$ is large enough (compared to the dipole atom-field
coupling $\Omega$, i.e, $\delta\gg\Omega$) to enable only virtual transitions
to occur between the states $|1\rangle$ and $|2\rangle$. In this regime, the
effective atom-field coupling parameter inside the cavity is $\chi=\Omega
^{2}/\delta$ \cite{Scully}. We suppose, for simplicity, that the atom-field
coupling is turned on (off) suddenly at the instant the atom enters (leaves)
the cavity region\textbf{, }such that $\chi=0$ when the atom is outside the
cavity. The time-dependent (TD) functions $\zeta(t)$ and $\xi(t)$ allow for
the parametric and linear amplifications, respectively. We consider the atom,
prepared at time $t_{0}$ by the first Ramsey zone in a $\left|  1\right\rangle
$,$\left|  2\right\rangle $ superposition, to reach $C$ at time $t_{1}$ and
leaves it at $t_{2}$. The linear and parametric pumping are supposed to be
turned on also at $t_{0}$ and turned off at a convenient time $t\geq t_{2}$.

The Schr\"{o}dinger state vector associated with Hamiltonian (\ref{Eq1}) can
be written using
\begin{equation}
|\Psi\left(  t\right)  \rangle=\operatorname*{e}\nolimits^{i\omega_{0}%
t/2}\left|  1\right\rangle \left|  \Phi_{1}\left(  t\right)  \right\rangle
+\operatorname*{e}\nolimits^{-i\omega_{0}t/2}\left|  2\right\rangle \left|
\Phi_{2}\left(  t\right)  \right\rangle \mathrm{{,}}\label{Eq2}%
\end{equation}
where $|\Phi_{\ell}\left(  t\right)  \rangle=\int\frac{d^{2}\alpha}{\pi
}\mathcal{A}_{\ell}\left(  \alpha,t\right)  |\alpha\rangle$, $\ell=1,2$, the
complex quantity $\alpha$ standing for the eigenvalues of $a$, and
$\mathcal{A}_{\ell}\left(  \alpha,t\right)  =\left\langle \alpha,\ell\left|
\Psi\left(  t\right)  \right.  \right\rangle $ are the expansion coefficients
for $|\Phi_{\ell}\left(  t\right)  \rangle$ in the basis of coherent-state,
$\left\{  |\alpha\rangle\right\}  $. Using the orthogonality of the atomic
states and Eqs. (\ref{Eq1}) and (\ref{Eq2}) we obtain the uncoupled TD
Schr\"{o}dinger equations:
\begin{equation}
i\frac{d}{dt}|\Phi_{\ell}\left(  t\right)  \rangle=\mathbf{H}_{\ell}%
|\Phi_{\ell}\left(  t\right)  \rangle\mathrm{{,}}\label{Eq3}%
\end{equation}%
\begin{equation}
\mathbf{H}_{\ell}=\omega_{\ell}(t)a^{\dagger}a+\zeta(t)a^{\dagger^{2}}%
+\zeta^{\ast}(t)a^{^{2}}+\xi(t)a^{\dagger}+\xi^{\ast}(t)a\mathrm{{,}%
}\label{Eq4}%
\end{equation}
with $\omega_{\ell}(t)=\left[  \omega+\left(  -1\right)  ^{\ell}\chi\right]
$. Note that the problem has been reduced to that of a cavity field under
parametric and linear amplifications, whose frequency $\omega$ is shifted by
$-\chi$ ($+\chi$) when interacting with the atomic state $\left|
1\right\rangle $ ($\left|  2\right\rangle $), during the time interval
$\tau=t_{2}-t_{1}$.

\textit{Time-dependent invariants}: To solve the Schr\"{o}dinger Eq.
(\ref{Eq3}) we employ the time-dependent invariants of Lewis and Riesenfeld
\cite{Lewis}. However, instead of proposing an invariant associated with the
Hamiltonian (\ref{Eq4}), we first perform a unitary transformation on Eq.
(\ref{Eq3}) in order to reduce it to a form which already has an associated
known invariant. Thus, under a unitary transformation represented by the
operator $S(\varepsilon_{\ell})$ ($\varepsilon_{\ell}$ standing for a set of
TD group parameters which may also depend on the atomic state $\ell$), we
obtain from Eq. (\ref{Eq3})
\begin{equation}
i\frac{d}{dt}\left|  \Phi_{\ell}^{S}(t)\right\rangle =\mathcal{H}_{\ell}%
^{S}\left|  \Phi_{\ell}^{S}\left(  t\right)  \right\rangle \mathrm{{,}%
}\label{Eq5}%
\end{equation}
where the transformed Hamiltonian and wave vector are given by
\begin{subequations}
\begin{align}
\mathcal{H}_{\ell}^{S}  & =S^{\dagger}(\varepsilon_{\ell})\mathcal{H}_{\ell
}S(\varepsilon_{\ell})+i\frac{dS^{\dagger}(\varepsilon_{\ell})}{dt}%
S(\varepsilon_{\ell})\mathrm{{,}}\label{Eq6a}\\
\left|  \Phi_{\ell}^{S}\left(  t\right)  \right\rangle  & =S^{\dagger
}(\varepsilon_{\ell})|\Phi_{\ell}\left(  t\right)  \rangle\mathrm{{.}%
}\label{Eq6b}%
\end{align}

In what follows we employ two theorems to obtain the solution of the TD
Schr\"{o}dinger Eq. (\ref{Eq3}): a) on the one hand, a theorem exposed in
\cite{Salomon} asserts that if $I_{\ell}(t)$ is an invariant associated to
$\mathcal{H}_{\ell}$ (i.e., $dI_{\ell}(t)/dt=\partial I_{\ell}/\partial
t+i\left[  \mathcal{H}_{\ell},I_{\ell}(t)\right]  =0$), then the transformed
operator $I_{\ell}^{S}(t)=S^{\dagger}(\varepsilon_{\ell})I_{\ell
}(t)S(\varepsilon_{\ell})$ becomes an invariant associated to $\mathcal{H}%
_{\ell}^{S}$; b) on the other hand, from Lewis and Riesenfeld's well-known
theorem \cite{Lewis}, it follows that a solution of the Schr\"{o}dinger
equation is an eigenstate of the Hermitian invariant $I_{\ell}(t)$ multiplied
by a TD phase factor. It follows from a) and b) that the solutions of Eq.
(\ref{Eq3}) are given by $|\Phi_{\ell}\left(  t\right)  \rangle=S(\varepsilon
_{\ell})\left|  \Phi_{\ell}^{S}\left(  t\right)  \right\rangle =S(\varepsilon
_{\ell})\exp\left[  i\phi_{\ell,m}^{S}(t)\right]  \left|  m,t\right\rangle
_{S}$ ($m=0,1,2,...$), where $\left|  m,t\right\rangle _{S}$ is the eigenstate
of the invariant \cite{Puri} and the Lewis and Riesenfeld phase \cite{Lewis}
obeys
\end{subequations}
\begin{equation}
\phi_{\ell,m}^{S}(t)=\int_{t_{i}}^{t}dt^{\prime}{}_{S}\left\langle
m,t^{\prime}\right|  \left(  i~\partial/\partial t^{\prime}-\mathcal{H}_{\ell
}^{S}\right)  \left|  m,t^{\prime}\right\rangle _{S}\mathrm{{.}}%
\label{EqPhase}%
\end{equation}
It is straightforward to verify that under the unitary transformation carried
out by the operator $S(\varepsilon_{\ell})$ the TD phase is invariant:
$\phi_{\ell,m}^{S}(t)=\phi_{\ell,m}(t)$.

\textit{The transformed Hamiltonian}: Next, we associate the unitary
transformation with the squeeze operator $S(\varepsilon_{\ell})=\exp\left[
\frac{1}{2}\left(  \varepsilon_{\ell}a^{\dagger^{2}}-\varepsilon_{\ell}^{\ast
}a^{2}\right)  \right]  $, where the complex TD function $\varepsilon_{\ell
}(t)=r_{\ell}(t)\operatorname*{e}\nolimits^{i\varphi_{\ell}(t)}$ includes the
squeeze parameters $r_{\ell}(t)$ and $\varphi_{\ell}(t)$. ($r_{\ell}(t)$ is
associated with a squeeze factor while $\varphi_{\ell}(t)$ defines the
squeezing direction in phase space.) Moreover, the TD parameters for the
parametric and linear amplifications are written as $\zeta(t)=\kappa
(t)\operatorname*{e}\nolimits^{i\eta(t)}$ and $\xi(t)=\varkappa
(t)\operatorname*{e}\nolimits^{i\varpi(t)}$, respectively. The squeeze
parameters ($r_{\ell}(t) $, $\varphi_{\ell}(t)$), the amplification amplitudes
($\kappa(t)$,$\varkappa(t)$) and frequencies ($\eta(t)$,$\varpi(t)$) are real
TD functions. From the above assumptions and after a lengthy calculation, the
transformed Hamiltonian becomes
\begin{equation}
\mathcal{H}_{\ell}^{S}=\Omega_{\ell}(t)a^{\dagger}a+\Lambda_{\ell
}(t)a^{\dagger}+\Lambda_{\ell}^{\ast}(t)a+\digamma_{\ell}(t)\mathrm{{,}%
}\label{Eq9}%
\end{equation}
provided that its TD coefficients satisfy
\begin{subequations}
\begin{align}
\Omega_{\ell}(t)  & =\omega_{\ell}(t)+2\kappa(t)\tanh r_{\ell}(t)\cos\left(
\eta(t)-\varphi_{\ell}(t)\right)  \mathrm{{,}}\label{Eq10a}\\
\Lambda_{\ell}(t)  & =\xi(t)\cosh r_{\ell}(t)+\xi^{\ast}(t)\operatorname*{e}%
\nolimits^{i\varphi_{\ell}(t)}\sinh r_{\ell}(t)\mathrm{{,}}\label{Eq10b}\\
\digamma_{\ell}(t)\mathrm{{}}  & =\kappa(t)\tanh r_{\ell}(t)\cos\left(
\eta(t)-\varphi_{\ell}(t)\right)  \mathrm{{,}}\label{Eq10c}%
\end{align}
while the squeeze parameters $r_{\ell}(t)$ and $\varphi_{\ell}(t)$ are
determined by solving the coupled differential equations
\end{subequations}
\begin{subequations}
\begin{align}
\overset{.}{r}_{\ell}(t)  & =2\kappa(t)\sin\left(  \eta(t)-\varphi_{\ell
}(t)\right)  \mathrm{{,}}\label{Eq11a}\\
\overset{.}{\varphi}_{\ell}(t)  & =-2\omega_{\ell}(t)-4\kappa(t)\coth\left(
2r_{\ell}(t)\right)  \cos\left(  \eta(t)-\varphi_{\ell}(t)\right)
\mathrm{{.}}\label{Eq11b}%
\end{align}
It is evident from these relations that the TD group parameters $\varepsilon
_{\ell}(t)$, defining the unitary operator $S(\varepsilon_{\ell})$, depends on
the atomic state $\ell$, as assumed from the beginning.

\textit{The evolution operators}: With the Hamiltonian (\ref{Eq9}) at hand we
return to the solution of the Schr\"{o}dinger Eq. (\ref{Eq5}). The Invariant
associated to this Hamiltonian is given by \cite{Puri}
\end{subequations}
\begin{equation}
I_{\ell}^{S}(t)=a^{\dagger}a-\theta_{\ell}(t)a^{\dagger}-\theta_{\ell}^{\ast
}(t)a+\mathfrak{f}_{\ell}(t)\mathrm{{,}}\label{EqInv}%
\end{equation}
$\theta_{\ell}(t)$ being a solution to the equation $i\overset{.}{\theta
}_{\ell}(t)=\Omega_{\ell}(t)\theta_{\ell}(t)+\Lambda_{\ell}(t)$ while
$\overset{.}{\mathfrak{f}}_{\ell}(t)=\theta_{\ell}^{\ast}(t)\Lambda_{\ell
}(t)-$ $\theta_{\ell}(t)\Lambda_{\ell}^{\ast}(t)=id\left|  \theta_{\ell
}(t)\right|  ^{2}/dt$. The application of the invariant method leads to the
wave vector \cite{Puri}%
\begin{equation}
\left|  \Phi_{\ell}^{S}(t)\right\rangle =\operatorname*{e}\nolimits^{i\phi
_{\ell,m}(t)}\left|  m,t\right\rangle _{S}=\operatorname*{e}\nolimits^{i\phi
_{\ell,m}(t)}D\left[  \theta_{\ell}(t)\right]  \left|  m\right\rangle
\ (m=0,1,2,...),\label{EqSV}%
\end{equation}
where $\left|  m\right\rangle $ is the number state and $D\left[  \theta
_{\ell}(t)\right]  =\exp\left[  \theta_{\ell}(t)a^{\dagger}-\theta_{\ell
}^{\ast}(t)a\right]  $ is the displacement operator.

Therefore, the solutions of the Schr\"{o}dinger Eq. (\ref{Eq3}), which form a
complete set, read $\left|  \Phi_{\ell}(t)\right\rangle =S\left[
\varepsilon_{\ell}(t)\right]  \left|  \Phi_{\ell}^{S}(t)\right\rangle
=U_{\ell}(t)\left|  m\right\rangle \mathrm{{,}}$ where $U_{\ell}%
(t)=\Upsilon_{\ell}(t)S\left[  \varepsilon_{\ell}(t)\right]  D\left[
\theta_{\ell}(y)\right]  R\left[  \Omega_{\ell}(t)\right]  $ is a unitary
operator composed, in addition to the squeeze and the displacement operators,
of a global phase factor $\Upsilon_{\ell}(t)=\exp\left\{  -\frac{i}{2}\left[
\beta(t)-\omega t\right]  \right\}  $ and the rotation operator (coming from
the TD Lewis and Riesenfeld phase factor) $R\left[  \Omega_{\ell}(t)\right]
=\exp\left[  -ia^{\dagger}a\beta_{\ell}(t)\right]  \mathrm{{,}}$ with
$\beta_{\ell}(t)=\int_{t_{i}}^{t}\Omega_{\ell}(t^{\prime})dt^{\prime}$. Hence,
for the solution of Schr\"{o}dinger Eq. (\ref{Eq3}), we find $\left|
\Phi_{\ell}(t)\right\rangle =U_{\ell}(t)U_{\ell}^{\dagger}(t_{i})\left|
\Phi_{\ell}(t_{i})\right\rangle \mathrm{{,}}$which finally defines the
evolution operators $\mathbb{U}_{\ell}(t,t_{i})=U_{\ell}(t)U_{\ell}^{\dagger
}(t_{i}). $

\textit{Evolution of the atom-field state}: Let us assume that the micromaser
cavity is prepared at time $t_{0}$ in a single-mode coherent state
$|\alpha\rangle$\ by a monochromatic source. As mentioned above, the linear
and parametric pumping are supposed to be turned on also at $t_{0}$, at the
time the atom is prepared by the first Ramsey zone in the superposition state
$c_{1}\left|  1\right\rangle +c_{2}\left|  2\right\rangle $. Evidently, the
evolution operators $\mathbb{U}(t_{1},t_{0})$ and $\mathbb{U}(t,t_{2})$
governing the dynamics of the cavity-field state while the atom is outside the
cavity, do not depend on the state of the two-level atom. However, the
operator $\mathbb{U}_{\ell}(t_{2},t_{1})$, given the evolution of the
cavity-field state during its interaction with the atom, does depend on the
atomic state and differs from the operators $\mathbb{U}(t_{1},t_{0})$ and
$\mathbb{U}(t,t_{2})$ only by the shifted frequency $\omega_{\ell}(t)$. With
this in mind it is straightforward to verify that the measurement of the
atomic state, after undergoing a $\pi/2$ pulse in the second Ramsey zone,
projects the cavity field in the ``Schr\"{o}dinger cat''-like state
\cite{Elsewhere}
\begin{equation}
|\Psi\left(  t\right)  \rangle=\mathcal{N}_{\pm}\left[  \pm\operatorname*{e}%
\nolimits^{i\omega_{0}t/2}c_{1}\mathsf{U}_{1}(t,t_{0})+\operatorname*{e}%
\nolimits^{-i\omega_{0}t/2}c_{2}\mathsf{U}_{2}(t,t_{0})\right]  \left|
\alpha\right\rangle \mathrm{{,}}\label{Eq16}%
\end{equation}
where the sign $+$ or $-$ occurs if the atom is detected in state $\left|
2\right\rangle $ or $\left|  1\right\rangle $, respectively, $\mathcal{N}%
_{\pm}$ accounts for the normalization factors, and the evolution operator
reads $\mathsf{U}_{\ell}(t,t_{0})=\mathbb{U}(t,t_{2})\mathbb{U}_{\ell}%
(t_{2},t_{1})\mathbb{U}(t_{1},t_{0})\mathrm{{.}}$ From Eq. (\ref{Eq16}) it
follows that, after measuring the atomic level used to generate the
superposition state of the radiation field, it is possible to control such
superposition by adjusting the TD driven parameters $\kappa(t)$,
$\varkappa(t)$, $\eta(t)$, and $\varpi(t)$.

\textit{Analytical solutions of the Characteristic equations} (\ref{Eq11a}%
,\ref{Eq11b}): Next, we investigate the situation where the cavity mode
$\left|  \alpha\right\rangle $ is resonant with the driven fields during the
time the atom is out of the cavity: from $t_{0}$ to $t_{1}$ and $t_{2}$ to
$t$. The parametric amplifier is assumed to operate in a degenerate mode in
which the \textit{signal} and the \textit{idler} frequencies coincide,
producing a single-mode driven field. In the resonant regime this single-mode
field has the same frequency $\omega$ as the cavity mode so that
$\eta(t)=-2\omega t$. For the resonant linear amplifier it follows that
$\varpi(t)=\omega t$. However, during the time interval the atom is inside the
cavity, from $t_{1}$ to $t_{2}$, it pulls the mode frequency out of resonance
with the driven fields establishing a dispersive regime of the amplification
process. In the resonant regime the solutions of the coupled differential Eqs.
(\ref{Eq11a},\ref{Eq11b}) are given by \cite{Salomon,Elsewhere}
\begin{subequations}
\begin{align}
\cosh\left(  2r(t)\right)   & =\sqrt{1+\mathcal{C}_{i}^{2}}\cosh\left[
\operatorname{arccosh}\left(  \frac{\cosh\left(  2r(t_{i})\right)  }%
{\sqrt{1+\mathcal{C}_{i}^{2}}}\right)  +u(t)\right]  \mathrm{{,}}%
\label{Eq18a}\\
\cos\left(  \varphi(t)-\eta(t)\right)   & =-\frac{\mathcal{C}_{i}}%
{\sqrt{\left(  1+\mathcal{C}_{i}^{2}\right)  \cosh^{2}\left(  u(t)-1\right)
}}\mathrm{{,}}\label{Eq18b}%
\end{align}
where $u(t)=4\int\kappa(t)dt$ and the constant of motion $\mathcal{C}%
_{i}\mathcal{=}\cos\left(  \varphi(t)-\eta(t)\right)  \sinh\left(
2r(t)\right)  $, depends on the initial values $r(t_{i})$, $\varphi(t_{i})$,
and $\eta(t_{i})$, where $i=0,2$. It is possible to show \cite{Elsewhere} that
in the dispersive regime Eqs. (\ref{Eq11a},\ref{Eq11b}) can be solved by
quadrature \cite{Salomon}, leading to a constant of motion, $\mathcal{C}%
_{1}=\cosh\left(  2r_{\ell}(t)\right)  +\mathfrak{P}_{\ell}\cos\left(
\varphi_{\ell}(t)-\eta(t)\right)  \sinh\left(  2r_{\ell}(t)\right)
\mathrm{{,}}$which now depends on the initial values $r(t_{1})$,
$\varphi(t_{1})$, and $\eta(t_{1})$. Despite the assumption that the
atom-field coupling is turned on (off) suddenly, these initial values must be
computed from the solutions for the resonant amplification regime at time
$t_{1}$. With this procedure we obtain the solutions for the resonant
amplification ($r(t_{1})$, $\varphi(t_{1})$) as a limit of those for the
dispersive amplification ($r(t_{1})$, $\varphi(t_{1})$) when $\chi
\rightarrow0$. The parameter $\mathfrak{P}_{\ell}=(-1)^{\ell}2\kappa/\chi$,
defined for a constant amplification amplitude $\kappa$, is an effective
macroscopic coupling. Therefore, for the dispersive regime we find three
different solutions depending on $\left|  \mathfrak{P}_{\ell}\right|  $: the
strong ($\left|  \mathfrak{P}\right|  >1$), the weak ($\left|  \mathfrak{P}%
_{\ell}\right|  <1$), and the critical coupling ($\left|  \mathfrak{P}_{\ell
}\right|  =1$). Considering the weak coupling regime, the TD squeeze
parameters when $\mathcal{C}_{1}>\sqrt{1-\mathfrak{P}_{\ell}^{2}}$ are given by%

\end{subequations}
\begin{subequations}
\begin{align}
\cosh\left(  2r_{\ell}(t)\right)   & =\frac{\mathcal{C}_{1}}{1-\mathfrak{P}%
_{\ell}^{2}}\left\{  1-\left|  \mathfrak{P}_{\ell}\right|  \frac
{\sqrt{\mathcal{C}_{1}^{2}-(1-\mathfrak{P}_{\ell}^{2})}}{\mathcal{C}_{1}%
}\right. \nonumber\\
& \left.  \times\sin\left[  \arcsin\left(  \frac{\mathcal{C}_{1}\left|
\mathfrak{P}_{\ell}\right|  }{\sqrt{\mathcal{C}_{1}^{2}-(1-\mathfrak{P}_{\ell
}^{2})}}\right)  -\frac{\kappa\sqrt{1-\mathfrak{P}_{\ell}^{2}}}{\left|
\mathfrak{P}_{\ell}\right|  }(t-t_{1})\right]  \right\}  ,\label{Eq20a}\\
\cos\left(  \varphi_{\ell}(t)-\eta(t)\right)   & =\pm\frac{\mathcal{C}%
_{1}-\cosh\left(  2r_{\ell}(t)\right)  }{\mathfrak{P}_{\ell}\left|
\sinh\left(  2r_{\ell}(t)\right)  \right|  }{.}\label{Eq20b}%
\end{align}

\textit{A protocol for engineering mesoscopic cavity-field states}: To prepare
a particular superposition state from (\ref{Eq16}) we follow a three-step
protocol. 1) First, we adjust the amplitude $\kappa$ of the parametric
amplification and the atom-field interaction time $\tau=t_{2}-t_{1}$ in order
to obtain a particular angle $\Theta/2$ $=\left(  \varphi_{1}(t_{2}%
)-\varphi_{2}(t_{2})\right)  /2$ defined by the squeezing directions of the
states composing the ``Schr\"{o}dinger cat''-like superposition. 2) Next, the
desired excitation of the prepared state can be achieved by manipulating the
excitation of the initial coherent state injected into the cavity and/or the
amplitude of the linear amplification (the strength of the parametric
amplification has been fixed in the first step) and/or the time interval of
the amplification process. 3) Finally, the amplitude of both states composing
the ``Schr\"{o}dinger cat''-like superposition can be adjusted through the
probability amplitudes of the atomic superposition state prepared in the first
Ramsey zone.

Evidently, the squeezed superposition in Eq. (\ref{Eq16}) was ideally
prepared. In a real engineering process the dissipative mechanisms of both the
cavity and the two-level atom, despite of the fluctuations intrinsic to their
interaction, must be taken into account. The complex calculations involved in
the engineering process of quantum states under the realistic quantum
dissipation and fluctuation conditions can be surpassed through the
phenomenological-operator approach as presented in Refs. \cite{Almeida,Serra}.
However, we will not consider in the present work the action of the reservoir
in the preparation of the squeezed superposition (\ref{Eq16}), since the time
interval required for this prepartion, of order of $10^{-4}$-$10^{-5}$ s, is
considerably smaller than the relaxation times of both the cavity field and
the two-level atom, around $\tau_{R}\approx10^{-2}$ s \cite{Brune1,Walther}.
Therefore, as usual for estimating the decoherence time, we next consider that
an ideally prepared state is submitted to action of a quantum reservoir. In
addition, we will be interested in the action of a vacuum-squeezed reservoir
at absolute zero whose density operator reads $\prod_{k}S_{k}\left|
0_{k}\right\rangle \left\langle 0_{k}\right|  S_{k}^{\dagger}$, $S_{k}$ being
the squeezed operator for the $k$th bath oscillator mode. We are here
considering that, somehow, it is possible to completely describe all the
mechanisms of dissipation of the cavity by the action of a vacuum-squeezed
reservoir. Describing the reservoir by a collection of harmonic oscillators
$\sum_{k}\hbar\omega_{k}b_{k}^{\dagger}b_{k}$ and its interaction with the
cavity mode by $\sum_{k}\hbar(\lambda_{k}a^{\dagger}b_{k}+\lambda_{k}^{\ast
}ab_{k}^{\dagger})$, \ the decoherence time deduced from the idempotency
defect of the reduced density operator of the cavity field, as suggested in
\cite{Piza}, is given by
\end{subequations}
\begin{equation}
\mathbf{\tau}=\frac{\mathbf{\tau}_{R}}{2\left|  (2N+1)\left(  \left\langle
a^{\dagger}\right\rangle \left\langle a\right\rangle -\left\langle a^{\dagger
}a\right\rangle \right)  +2\operatorname{Re}\left[  M\left(  \left\langle
a^{\dagger}\right\rangle ^{2}-\left\langle \left(  a^{\dagger}\right)
^{2}\right\rangle \right)  \right]  -N\right|  },\label{Eq21}%
\end{equation}
where $\mathbf{\tau}_{R}$ is the relaxation time, $N=\sinh^{2}(\widetilde{r}%
)$, and $M=-\operatorname{e}^{i\widetilde{\varphi}}\sinh(2\widetilde{r})/2$,
$\widetilde{r}$ and $\widetilde{\varphi}$ being the squeeze parameters of the
vacuum reservoir \cite{Elsewhere}. The mean values are computed from the
prepared squeezed superposition (\ref{Eq16}). Since the excitation of the
initial coherent state $\alpha$ and the squeeze parameters ($r(t_{2}%
),\varphi_{\ell}(t_{2}) $) have been fixed by the engineering protocol, we
note that Eq. (\ref{Eq21}) depends only on the reservoir squeeze parameters
($\widetilde{r}$,$\widetilde{\varphi}$). Considering the situation where
$\alpha$ is real and $\exp(-2\alpha^{2})\approx0$ (implying $\alpha
\gtrapprox\sqrt{2}$), the maximization of the decoherence time $\mathbf{\tau}$
with respect to these parameters leads to the results
\begin{subequations}
\begin{align}
\widetilde{r}_{A}  & \mathbf{=}r+\ln(1+4\alpha^{2})/4,\quad\widetilde{\varphi
}_{A}=0,\label{Eq22a}\\
\widetilde{r}_{B}  & =r-\ln(1+4\alpha^{2})/4,\quad\widetilde{\varphi}_{B}%
=\pi,\label{Eq22b}%
\end{align}
when fixing $\Theta=2n\pi$ ($n$ integer), i.e.,\ the states composing the
superposition (\ref{Eq16}) are squeezed in the same direction. When
$\Theta\neq2n\pi$, the maximum of $\mathbf{\tau}$ turns out to be smaller than
that for $\Theta=2n\pi$, given either by the pair ($\widetilde{r}_{A}$,
$\widetilde{\varphi}_{A}$) or ($\widetilde{r}_{B}$, $\widetilde{\varphi}_{B}
$) when considering $\varphi_{\ell}(t_{2})=$ $(2m_{\ell}+1)\pi$ or
$\varphi_{\ell}(t_{2})=2m_{\ell}\pi$ ($m_{\ell}$ integer), respectively (note
that $n=\left|  m_{1}-m_{2}\right|  $). Observe that the direction of
squeezing of both states composing the superposition (\ref{Eq16}), defined by
the angle $\varphi_{1}(t_{2})$ or $\varphi_{2}(t_{2})$ has to be perpendicular
to the direction of squeezing of the vacuum reservoir.

Next, we compute the ``distance'' in phase space between the centers of the
quasi-probability distribution of the individual states composing the prepared
superposition (\ref{Eq16}). This distance is defined by the quadratures of the
cavity field $X=(a^{\dagger}+a)/2$ and $Y=(a-a^{\dagger})/2i$, as $D=\left[
\left(  \left\langle X\right\rangle _{2}-\left\langle X\right\rangle
_{1}\right)  ^{2}+\left(  \left\langle Y\right\rangle _{2}-\left\langle
Y\right\rangle _{1}\right)  ^{2}\right]  ^{1/2}$, the subscripts $1$,$2$
referring to the two states composing the superposition. When considering
$\varphi_{1}(t_{2})=$ $\varphi_{2}(t_{2})=(2m+1)\pi$ or $2m\pi$, the distance
becomes $D=\left\langle X\right\rangle _{2}-\left\langle X\right\rangle
_{1}=2\alpha\exp(r)$ or $2\alpha\exp(-r) $, respectively. We will focus on the
case $\varphi_{1}(t_{2})=$ $(2m_{1}+1)\pi$, since it results in a large
distance $D$ between the two states composing what we actually want to be a
mesoscopic superposition. Assuming the squeezing factor $r$ so that
$\exp(-2r)\approx$ $0$ in addition to $\exp(-2\alpha^{2})\approx$ $0$, the
decoherence time and the mean photon number of the prepared state, following
from the values ($\widetilde{r}_{A}$, $\widetilde{\varphi}_{A}$), reads
\end{subequations}
\begin{equation}
\mathbf{\tau}\approx\mathbf{\tau}_{R}/\alpha,\quad\left\langle n\right\rangle
=\left\langle a^{\dagger}a\right\rangle \approx\alpha^{2}\exp(2r).\label{Eq23}%
\end{equation}
Remarkably, with the approximations $\exp(-2r)$, $\exp(-2\alpha^{2})\approx0$,
the decoherence time for the prepared cavity-field state when $\varphi
_{1}(t_{2})=$ $\varphi_{2}(t_{2})=(2m_{2}+1)\pi$ --- \textit{under the action
of a vacuum reservoir squeezed in the direction }$\widetilde{\varphi}_{A}=0$
--- turns out to be practically independent of the parameter $r$ and thus of
their own intensity $\left\langle n\right\rangle $ and distance $D$. From the
result in Eq. (\ref{Eq23}) we conclude that it is convenient to start from a
coherent state $\alpha$ as small as possible (within the limit $\exp
(-2\alpha^{2})\approx$ $0$) and to adjust the macroscopic coupling parameter
$\left|  \mathfrak{P}_{\ell}\right|  $ in order to obtain a large squeeze
factor and so a large intensity of the prepared state, since we are actually
interested in mesoscopic superpositions. We stress that even considering the
weak coupling regime ($\left|  \mathfrak{P}_{\ell}\right|  <1$) we obtain,
from Eqs.(\ref{Eq18a}) and (\ref{Eq20a}), large squeeze parameters.
Considering $\left|  \mathfrak{P}_{\ell}\right|  =0.1$, $\alpha=\sqrt{2}$, and
the experimental running time about $\ 2\times10^{-4}s$, we get a
superposition state where $r\approx2$ and $\left\langle n\right\rangle
\approx10^{2}$ photons.

The mechanism behind this result is the degree of entanglement between the
prepared state and the modes of the reservoir, which depends on the relative
direction of their squeezing, defined by the angles $\varphi_{1}(t_{2}%
)=$\textit{\ }$\varphi_{2}(t_{2})$ and $\widetilde{\varphi}_{A}$. A result
supporting this argument is presented in \cite{Knight} where it is shown that
the injection of two modes, squeezed in perpendicular directions, in a $50/50$
beam splitter does not generate an entangled state. A careful analysis of the
dependence on the degree of entanglement and the relative direction of
squeezing between a prepared state and its multimode reservoir --- a
collection of independent beam splitters --- is presented in \cite{Elsewhere}.
Despite the fact that the mechanism behind the long-lived mesoscopic
superpositions is mainly the perpendicular directions between the squeezing of
the prepared state and the reservoir modes, the magnitude of the parameter $r$
plays a crucial role in the present scheme for producing the mesoscopic
superposition by increasing both their intensity $\left\langle n\right\rangle
$ and distance $D$ in phase space.

The values presented above for $\mathbf{\tau}$, $\left\langle n\right\rangle
$, and $D$ are to be compared with those when considering a non-squeezed
($NS$) cavity-field state ($\left\langle n\right\rangle _{NS}=\alpha^{2}$,
$D_{NS}=2\alpha$ ) under the influence of $i)$ a squeezed reservoir, resulting
in the decoherence time $\mathbf{\tau}_{i}\approx\mathbf{\tau}_{R}/\alpha$,
and $ii)$ a non-squeezed reservoir, such that $\mathbf{\tau}_{ii}%
\approx\mathbf{\tau}_{R}/2\alpha^{2}$. Note that in both cases $i)$ and $ii)$
we obtain the rates $\left\langle n\right\rangle /\left\langle n\right\rangle
_{NS}\approx\exp(2r)$ and $D/D_{NS}\approx\exp(r)$. Therefore, despite the
exponential increase in the rates for both excitation and distance we still
get $\mathbf{\tau\approx\tau}_{i}$ when comparing our results with previous
schemes in the literature, where a squeezed reservoir is assumed for the
enhancement of the decoherence time \cite{Kim}; for non-squeezed cavity-field
states and reservoir, we obtain a still better result $\mathbf{\tau
\approx\alpha\tau}_{ii}$.

It is interesting to note that we could have separated the process of
Hamiltonian (\ref{Eq1}) into two successive simpler processes: first creating
the superposition state (with the first three terms of Hamiltonian
(\ref{Eq1})) and then applying the parametric amplification (forth and fifth
term of (\ref{Eq1}), without need of linear amplification). When considering
the whole process simultaneously, as done in this work, the squeezing
directions in phase space of both states composing the superposition can be
adjusted independently. Evidently, this is not necessary for the proposal
presented in this paper, where the components of the superposition have to be
squeezed in the same direction. However, the possibility of squeezing the
components of a superposition state in different directions can be considered
for other applications as for state engineering in cavity QED \cite{Elsewhere}%
. In this connection, the linear amplification process (sixth and seventh
terms of Eq. (\ref{Eq1})) can be employed to achieve a higher excitation of
the engineered state. Moreover, when considering the whole process
simultaneously, we decrease the time interval of the experiment,\ minimizing
the noise effects coming from field and atomic decays, thus achieving a higher
fidelity for the generated squeezed superposition state. Even that we have not
computed the noise effects during the preparation of the squeezed
superposition state, it is always desirable to minimize the time interval of
the engineering process in order to maximize the fidelity of the prepared state.

The experimental implementation of the proposed scheme rely on the possibility
of engineering a squeezed reservoir as well as of parametrically driving
cavity-field radiation. We stress that a scheme to realize physically a
squeezed bath for cavity modes, via quantum-nondemolition-mediated feedback,
was already presented in Ref. \cite{Vitali}. However, the feedback process in
\cite{Vitali} does not eliminate the standard nonsqueezed bath and, as we have
stressed, our scheme requires a resulting optimal squeezed-vacuum reservoir.
The subject of quantum reservoir engineering has attracted attention specially
in the domain of trapped ions \cite{Poyatos,Matos} and, specifically, a scheme
has been presented for engineering squeezed-bath-type interaction for
protecting a two-level system against decoherence \cite{Lutkenhaus}.

Regarding parametric amplification of cavity fields, a technique was recently
suggested based on pulsed excitation of semiconductor layers (on the cavity
walls) by laser radiation \cite{Carugno}. Moreover, a proposal to implement
the parametric amplification of an arbitrary radiation-field state previously
prepared in a high-$Q$ cavity is presented in Ref. \cite{Parametric}. In this
work, the nonlinear process is accomplished through the dispersive
interactions of a three-level atom simultaneously with a classical driven
field and a previously prepared cavity mode whose state is supposed to be
squeezed. It is worth mention that all the treatment developed above in the
context of cavity quantum electrodynamics, for delaying the decoherence
process of a squeezed superposition by coupling it to a vacuum-squeezed
reservoir, can also be implemented in ion traps. We finally mention that the
proposal here presented might provide a motivation for future theoretical and
experimental investigations.

\begin{acknowledgments}
We wish to express thanks for the support from FAPESP (under contracts
\#99/11617-0, \#00/15084-5, and \#02/02633-6) and CNPq (Instituto do
Mil\^{e}nio de Informa\c{c}\~{a}o Qu\^{a}ntica), Brazilian agencies. We also
thank S. S. Mizrahi, R. Napolitano, V. V. Dodonov, and B. Baseia for helpful discussions.
\end{acknowledgments}

\end{document}